\documentstyle[12pt,epsf]{article}

\def\bi{\bigskip}
\def\noi{\noindent}
\def\be{\begin{equation}}
\def\en{\end{equation}}
\def\bq{\begin{eqnarray}}
\def\eq{\end{eqnarray}}
\def\bc{\begin{center}}
\def\ec{\end{center}}
\begin{document}
\rightline{\bf IFUG HEP-9901}
\begin{center}
{\large{\bf The $\omega \to \rho\pi$ transition and $\omega \to 3\pi$ decay
\footnote{Work supported by CONACyT under contract 397988-E}}}
\\[1.5cm]
\end{center}
\begin{center}
{\bf J.L. Lucio M.$^{(1)}$, M. Napsuciale$^{(1)}$, M.D. 
Scadron$^{(2)}$ and V.M. Villanueva$^{(1)}$.}\\[.3cm]
{\it $^{(1)}$ Instituto de F\'\i sica, Universidad de Guanajuato}\\
{\it Lomas del Bosque \# 103, Lomas del Campestre}\\
{\it 37150 Le\'on, Guanajuato; M\'exico}\\
\vspace{.5cm}

{\it $^{(2)}$ Department of Physics, University of Arizona}\\
{\it Tucson, AZ 85721, U.S.A.}
\end{center}

\noi PACS number(s): 13.25.-k, 13.25.Jx, 14.40.Cs

\vspace{2.5cm}

\begin{center}
{\bf Abstract.} 
\end{center}

We evaluate the $\omega \to \rho \pi$ transition and the
$\omega \to 3\pi$ decay using a Quark Level Linear Sigma Model
($QL\sigma M$).
We obtain $g^{QL\sigma M}_{\omega\rho\pi}= (10.33 - 14.75)$ $GeV^{-1}$
to be compared with other model dependent estimates averaging to
$g_{\omega\rho\pi} = 16 ~ GeV^{-1}$.
We show that in the $QL\sigma M$ a contact term is generated
for the $\omega \to 3 \pi$ decay. Although the contact contribution by
itself is small, the interference effects turn out to be important.
\setlength{\baselineskip}{1\baselineskip}

\newpage

\noi {\bf I. Introduction.} 

\bi

The problem of understanding low energy hadron dynamics is being supported by
recent experimental data obtained at Novosibirsk-VEPP-2M detector and by the 
DA$\Phi$NE facility which will soon produce large amounts of experimental 
data in 
the energy region around 1 GeV. This region is particularly interesting since  
the nonperturbative QCD effects that govern hadron dynamics are far from 
being completely understood.

\bi

This paper is concerned with the $\omega\to\rho\pi$ vertex and 
$\omega\to\pi\pi\pi$ decay. Existing experimental data seems to confirm the 
Gell-Mann-Sharp-Wagner (GSW) suggestion [1] that the $\omega\to\pi\pi\pi$ 
decay is dominated by the $\omega\rho\pi$ transition followed by the 
$\rho\to\pi\pi$ decay, although a $\omega\pi\pi\pi$ contact contribution can
not be excluded. The theoretical description of these problems have been 
considered by a number of authors using as different techniques  as 
approximate SU(3) symmetry[2], Vector Meson Dominance (VMD)[3], QCD sum 
rules [4,5,6] and effective chiral lagrangians [7].

\bi

It is our purpose in this paper to consider the Quark Level Linear Sigma 
Model (QL$\sigma$M) predictions for the $\omega \to \rho \pi$ transition and the 
$\omega\to\pi\pi\pi$ decay. The model describes the U(2)$\times$U(2) chiral
invariant interactions of effective quarks with  
pseudoscalar and scalar mesons. Vector mesons
are incorporated in the model as gauge bosons, even though in our work they 
only appear as external particles. In this model, besides the GSW mechanism,
the $\omega\to\pi\pi\pi$ decay proceeds through a quark box diagrams with the
$\omega$ and three pions in the box vertices, which can be interpreted as a 
contact term. By itself the contact term is not important: it leads to a 
$\Gamma (\omega\to\pi\pi\pi ) \cong$ 0.1 MeV. However its interference with 
the amplitude arising from the  GSW mechanism leads to a sizable 25\% effect 
in the decay rate.

\bi

The paper is organized as follows: In section II we introduce the model and 
discuss the determination of the coupling constants entering in the Lagrangian.
Then in section III we compute -working in the soft momentum limit for what 
follows- the QL$\sigma$M quark loop for $g_{\omega\rho\pi}$. In 
section IV we present calculations for  $\omega\to\pi\pi\pi$ within 
QL$\sigma$M. This include the GSW mechanism  and  quark box contributions. 
Comparison 
with the observed $\omega\to\pi\pi\pi$ decay rate is then made. In section V 
we draw our conclusions.

\bi

\bi

\noi {\bf II. The Model.}

\bi

The quark level L$\sigma$M describes the U(2)$\times$U(2) chiral invariant 
interaction of mesons and effective quarks. 
 We have
chosen to work with a pseudoscalar rather than a derivative coupling which 
has the advantage that no anomalous interactions 
are required to describe one pion processes. Vector mesons are  
incorporated in the model as gauge bosons, even though in our work they only 
appear as external particles. The $QL\sigma M$ Lagrangian is:  
\be
 {\cal L}_{int} =  \bar\psi \bigl[i\not D-M + g(S+i\gamma_5 P)\bigr] \psi + 
 (D_\mu B D^\mu B^\dagger )/2 +...
\en
 where  $B\equiv S+iP$ with $S,P$  scalar and pseudoscalar fields respectively 
($ P={1 \over \sqrt{2}}(\eta_0 + \vec\tau\cdot\vec\pi)$, scalar fields being 
defined in a similar way) and $\psi$ denotes the quark isospinor. Vector 
fields are introduced as gauge fields through the covariant derivative:

\bq
D_\mu\psi \equiv (\partial_\mu + ig_V V_\mu)\psi ,  \nonumber  \\
D_\mu B \equiv \partial_\mu B +ig_V[V_\mu,B].  
\eq
The quark mass matrix  $M$ in Eq.(1) is generated by spontaneous breaking 
of chiral symmetry, quark masses being related to the $ \pi qq$ coupling $g$ 
through the Goldberger-Trieman relation (GTR) $m_q =g f_\pi$. The ellipsis in 
Eq.(1) refers to vector meson kinetic and mass terms and to  
scalar-pseudoscalar Yukawa interactions which are not relevant for the  
purposes of this work.

The interaction  
terms involved in  the calculations presented in this paper are: 
\be
{\cal L}_{int}=g_V(\vec\rho \times \vec\pi ) \cdot \partial \vec\pi 
- \frac{g_V}{2}\bar\psi \gamma^\mu \vec\tau \psi \cdot \vec\rho_\mu 
- \frac{g_V}{2} \bar\psi \gamma^\mu \psi \omega_\mu.
\en

Thus, the model relates  the $\rho qq$ 
and $\omega qq$ couplings (${g_V\over 2}$) to the $\rho\pi\pi$ coupling 
($g_V$). This constant can be estimated from experimental data on  the 
$\rho\to\pi\pi$ decay.
Another determination comes from  Vector Meson Dominance as applied to the 
constituent quark level. The conventional VMD procedure leads to  the 
following 
relations between the $\rho -\gamma ~~(f_{\rho\gamma})$, $\omega -\gamma ~~ 
(f_{\omega\gamma})$, and the $ Vqq$ couplings:

\be
g_{\rho qq} = \frac{e m^2_\rho}{f_{\rho\gamma}}, \hspace{1.5cm}
g_{\omega qq} = \frac{em^2_\omega}{3 f_{\omega\gamma}} .
\en

The $\frac{1}{3}$ factor in the last expression arises due to the fact that
the isoscalar contribution to the quark electric charge is proportional to the quark 
baryonic number. Using the data [8] for the leptonic decays $\rho\to e^+e^-$, 
$\omega\to e+e^-$ we obtain $g^{\rho ll}_V\equiv g_{\rho qq} = 5.03$ and 
$g^{\omega ll}_V \equiv g_{\omega qq}=5.68$, whereas the $\rho \to \pi\pi$ 
decay leads to $g^{\rho\pi\pi}_V = 6.01$ (superindices in $g_V$ indicates the 
process from which it is extracted). 

\noi Summarizing, for $g_V$ in Eq.(2,3) we can use either 
$g^{\rho\pi\pi}_V = 6.01$, $g^{\rho ll}_V\equiv g_{\rho qq} = 5.03$ or  
$g^{\omega ll}_V \equiv g_{\omega qq}=5.68$. In the rest of the paper we 
report numerical results for these three different values of $g_V$, even 
though 
we argue below that the more accurate determination comes from 
$g^{\omega ll}_V$ (besides being close to the average value of 
$g^{\rho\pi\pi}_V$ and $g^{\rho ll}_V $).

\bi

\noi {\bf III. QL$\sigma$M and the $\omega\to \rho\pi$ transition.}

\bi

It is conventional to define the $g_{\omega\rho\pi}$ coupling in terms of the 
amplitude:

\be
M_{\omega\rho\pi} = g_{\omega\rho\pi} \epsilon_{\mu\nu\alpha\beta} P^k 
P^{\prime\nu} \epsilon^\alpha (\omega ) \eta^\beta (\rho ),
\en

\noi where $P(P^\prime )$ denote the $\omega (\rho )$ momentum and 
$\epsilon(\eta )$ are the respective polarization vectors. Although there is 
no phase space from which to measure an $\omega\to\rho\pi$ transition, this 
vertex can be extracted from many theoretical models. Very approximate SU(3) 
symmetry of the 1970's suggest [2] $g_{\omega\rho\pi} \approx 16~ GeV^{-1}$, 
while  
QCD sum rules obtain [4] 
$g_{\omega\rho\pi} \approx$ (15 $\to$ 17) $GeV^{-1}$, and the analog light 
cone sum rules method extracts [6] $g_{\omega\rho\pi} =15~ GeV^{-1}$. 
Recently QCD sum rules for the polarization operator in an external field 
concludes [5] $g_{\omega\rho\pi} \approx 16~ GeV^{-1}$.
It is interesting to mention that this kind of coupling affects strongly
the calculations for some other processes [9].

\bi

In the QL$\sigma$M the $\omega\rho\pi$ vertex is described in terms of the
quark loops of Fig (1). A straightforward calculations yields:

\[ {\cal M} \left[ \omega (Q, \eta )\to \rho (k, \varepsilon )+ \pi (r)
\right] = g_{\omega\rho\pi} \epsilon (k, r,\eta , \varepsilon ), \]

\noi where
\be
 g_{\omega\rho\pi} =-2i N_c m_q  g^2_V g~ (I^a+I^b),  
\en
\noi with
\be
 I^a \equiv \int \frac{d^4 \ell}{(2\pi )^4} \cdot \frac{1}{\nabla (\ell)
\nabla (\ell -k) \nabla (\ell -k-r) },
\en
\[ I^b = I^a (r \leftrightarrow k )\] 

\noi where $\nabla (p)\equiv p^2-m^2_q$.
Since we are interested in the $\omega\rho\pi$ vertex involved in 
$\omega\to\pi\pi\pi$ decay, we calculate the integrals in Eq. (7) in the soft 
pion momentum limit, $k,r\to 0$ . This amounts to take the 
leading term in an expansion of the amplitude in the external momenta $k$ and
$r$. In this approximation $I^a = I^b$ and we get for the $\omega\rho\pi$
 coupling:

\be
g^{QL\sigma M}_{\omega\rho\pi} = \frac{g^2_V N_c}{8 \pi^2 f_\pi}. 
\en

 Using the 
values of the coupling constants as previously estimated we conclude:

\be
 g^{QL\sigma M}_{\omega\rho\pi} = \left( 10.33,~13.19,~ 14.75 \right)GeV^{-1} .
\en

\noi These values are obtained using in Eq.(8) $g_V=(g^{\rho ll}_V, 
g^{\omega ll}_V, g^{\rho\pi\pi}_V)$ respectively. Thus, $QL\sigma M$ 
predictions for $g_{\omega\rho\pi}$ are slightly smaller than other 
determinations [2-7].

\bi

As a byproduct of our analysis we report the predictions of the $QL\sigma M$ 
for $\omega \to\pi^0 \gamma$ and $\rho^0\to\pi^0\gamma$ which 
are very similar to 
the ones presented so far, the only 
difference being the appearance of the $\gamma qq$ coupling instead of the 
$Vqq$ coupling. We obtain:

\be
{\cal M} (V(Q,\eta) \to \pi (r)\gamma (k,\varepsilon )= 
g^{QL\sigma M}_{V\pi\gamma} ~ \epsilon (k,r,\eta,\varepsilon ),
\en 
\noi where
\bq
g^{QL\sigma M}_{\rho\pi\gamma}&=& -i \frac{g_V e}{8\pi^2} \frac{g}{m_q} N_c
(e_u + e_d) \, , \nonumber \\ 
g^{QL\sigma M}_{\omega\pi\gamma}  &=& -i \frac{g_Ve}{8\pi^2} 
\frac{g}{m_q} N_c (e_u -e_d) \, .   \\
\nonumber 
\eq

\noi A worth noticing point is that both $g_{\omega\rho\pi}$ and 
$g_{V\pi\gamma}$ as calculated in this work, agree with those derived from 
a chiral lagrangian with vector mesons in the hidden scheme [10]. 

Using the GTR and  $g_V= g^{\rho ll}_V,~g^{\omega ll}_V,~g^{\rho\pi\pi}_V$ we 
obtain respectively: 

\bq
| g^{QL\sigma M}_{\omega\pi\gamma} | &=& (0.622,~0.703,~0.743) 
GeV^{-1}, \nonumber \\
| g^{QL\sigma M}_{\rho\pi\gamma} | &=& (0.207,~ 0.234,~ 0.247) 
GeV^{-1}. 
\eq

\noi These numbers are to be compared with the experimental 
results $|g^{exp}_{\omega\pi\gamma}| = 0.703 \pm 0.020~ GeV^{-1}$, 
$|g^{exp}_{\rho\pi\gamma}| = 0.29 \pm 0.037~ GeV^{-1}$.

Notice that the $QL\sigma M$ predictions for the $\omega \to \pi \gamma$ 
transition are in 
good agreement with the experimental data when $g_V=g^{\omega ll}_V$ - as 
extracted from the $\omega$ leptonic width - is used. On the other hand 
predictions for $\rho \to \pi\gamma$ agree with  experimental results within 
two standard deviations. These results and the fact that the measurements for 
the $\omega \to \pi\gamma$ decay rate are more accurate than for 
$\rho \to \pi\gamma$ lead us to consider $g_V=5.68$  as the most confident 
value for $g_V$.

\bi
\bi

\noi {\bf IV. QL$\sigma$M AND THE $\omega\to\pi\pi\pi$ DECAY}.

\bi

In this section we work out the QL$\sigma$M predictions for the 
$\omega\to\pi\pi\pi$ decay. Two mechanisms contribute to this process within
the model. The first one is through an intermediate $\rho$ in the $s,t, u$ 
channels as depicted in Fig(2), which involve the previously calculated 
$g_{\omega\rho\pi}$ and $g_{\rho\pi\pi}$. The second mechanism involves quark
boxes, as shown in Fig (3).

\bi

The $\rho$-mediated contribution leads to a $\omega\to\pi\pi\pi$ amplitude:

\be
 {\cal M}^{GSW} \left[ \omega (Q,\eta )\to \pi^+ (q) \pi^- (p)\pi^0 (r)
\right] = A^{GSW} ~\varepsilon (\eta, p, q, r), 
\en

\noi where:

\be
 A^{GSW} =2 g_{\omega\rho\pi} ~g_V ~\Big( \frac{1}{s-m^2_\rho} +
\frac{1}{t-m^2_\rho}+\frac{1}{u-m^2_\rho} \Big). 
\en

The amplitude corresponding to the quark boxes contribution is:

\be
 A^{box} = - 4 g_V~ g^3~ m_q~ I~ \varepsilon (\eta, p, q, r) 
\en 

\noi where

\[ I = \sum^{6}_{i=1} I^i, \]

\noi and

\be
 I^1 =\int \frac{d^4 \ell}{(2\pi )^4} \cdot \left[ \nabla (\ell ) 
\nabla (\ell -r)\nabla (\ell -r-p) \nabla (\ell - r -p-q) \right]^{-1}.
\en

\bi

The 5 remaining integrals are obtained from $I^1$ by permutations in Eq. (16) 
of the pion momenta $(p,q,r)$. We again consider the leading term 
of an expansion of this integral in the pions momenta, since higher order 
terms are expected to be suppressed by powers of 
$\Big(\frac{m_\pi}{m_q}\Big)^2$. In 
this approximation we obtain:

\be
 A^{box} =-\frac{g_V}{4\pi^2 f^3_\pi}. 
\en

The decay rate $\Gamma (\omega \to\pi\pi\pi )$ is given by:

\be
 \Gamma  (\omega \to 3\pi )= \frac{g^2_V~g^2_{\omega\rho\pi} 
m^3_\omega}{768 \pi^3} ~ J, 
\en

\noi where $J$ stands for the phase space integral

\be
 J= \int^{(1-\beta)^2}_{4\beta} dx \int^{y_-}_{y_+} dy |f(x,y)|^2 \Delta
(x,y), 
\en

\noi with 

\be
 y_\pm = \frac{1}{2} (1+3\beta -x \mp\sqrt{(1-\frac{4\beta}{x}) (x-(1+
\sqrt{\beta})^2) (x-(1-\sqrt{\beta}))^2}, 
\en

\noi and 

\[f(x,y)=f^{GSW} (x,y) + f^{box} (x,y) \] \, ,

\noi where

\bq
f^{GSW} (x,y) &=& \frac{1}{x-\alpha}+\frac{1}{y-\alpha} + 
\frac{1}{1+3\beta -x-y-\alpha}, \nonumber \\
f^{box} (x,y) &=& -\frac{ m^2_\omega}{24 \pi^2 f^3_\pi 
g_{\omega\rho\pi}}. \nonumber
\eq 

The Kibble determinant $\Delta$ in Eq.(19) is given by:

\be
 \Delta (x,y) =xy (1+3 \beta )-x^2 y - xy^2 -\beta (1-\beta )^2 ,
\en 

\noi where

\[ x=\frac{s}{m^2_\omega} , ~~~~~ y= \frac{t}{m^2_\omega} , ~~~~
\beta =\frac{m^2_\pi}{m^2_\omega},~~~~ \alpha =
\frac{m^2_\rho}{m^2_\omega}. \]

The phase space integral has been worked out by Thews [3] in the case of a 
constant matrix element. Since the numerical evaluation of the integral is 
straightforward, below we report the results according to the exact numerical 
evaluation, which of course reproduce the results of Ref. [3] in the case 
of a constant matrix element.

\bi

The quark box contribution to the $\omega\to\pi\pi\pi$ decay rate is small:

\be
 \Gamma_{box} (\omega \to 3\pi ) \simeq 0.11~ MeV. 
\en

In table 3 we summarize the numerical results for the $\omega\to 3\pi$ decay 
width for the three possible values of $g_V$.

\bi

\begin{center}
{\bf Table I}
\end {center}
\bi
\begin{tabular}{|c|c|c|c|} \hline
$g_\rho$ & $g_{\omega\rho\pi}(GeV^{-1})$ & $\Gamma^{GSW} (\omega\to 3\pi) 
(MeV)$ & $\Gamma^{TOT} (\omega\to 3\pi )(MeV)$ \\ \hline
5.03   & 10.33            & 2.72                  & 3.79 \\      \hline
5.68   & 13.19            & 5.66                  & 7.39 \\      \hline
6.01   & 14.75            & 7.92                  & 10.06 \\  \hline
\end{tabular}

\bi
These results have to be compared with the experimental data:

\be
 \Gamma_{exp} (\omega \to 3 \pi ) = 7.5 \pm 0.1 MeV 
\en

Notice that while the quark box contribution alone is negligible, the 
interference with the $\rho$ mediated amplitude is important.

In the $QL\sigma M$ the non-resonant 
contribution arising from quark box diagrams (Fig.3) necessarily exist.
From the numerical results reported in Table 3 we conclude that $QL\sigma M$ 
predictions are in agreement with experimental data within the $25\%$ 
uncertainty associated with the determination of $g_V$. It is worth remarking 
that the most clean extraction of this coupling (from $\omega$ leptonic width) 
is close to the experimental result.
\bi
\bi

\noi {\bf V. SUMMARY.}

\bi

We worked out the quark level Linear Sigma Model predictions for the
$\omega\rho\pi$ coupling constant. We obtain $g^{QL\sigma M}_{\omega\rho\pi}=
(10.33,~ 13.19,~14.75) GeV^{-1}$. These values correspond to $g_V$ as extracted 
from $\rho$ leptonic width, $\omega$ leptonic width and $\rho\to\pi\pi$ 
respectively. The $QL\sigma M$ predictions for $g_{\omega\rho\pi}$
are smaller than other determinations [2-7].

\bi

The non-resonant contribution to the $\omega \to 3\pi$ decay naturally arise 
within the $QL\sigma M$.
We evaluated the boxes contributions (Fig. (3)) to the $\omega\to 3\pi$
decay. We find an amplitude which leads to a small decay rate by itself. 
However,
the interference of the boxes contribution with the $\rho$ mediated amplitude lead to a
sizable ($25- 30 \% $) effect in the decay rate.

The amplitudes for the $\omega \to \pi \gamma ~ (g_{\omega\pi\gamma})$  and
$\rho \to\pi\gamma ~(g_{\rho\pi\gamma})$ can be obtained as a byproduct
of the $g_{\omega\rho\pi}$ calculation. The agreement with 
experimental data is good in the case of the $\omega \to\pi\gamma $ decay. In the case of the
$\rho \to\pi\gamma$ decay; QL$\sigma$M predictions agree with experimental data within two 
standard deviations.

\bi

{\bf Acknowledgments.}

\bi
One of us (M.N.) wishes to thank the hospitality of the high energy 
physics theory group at UCSD where part of this work was done, his work was 
supported in part by a Conacyt grant $\#933053$ . Also the coauthor MDS 
appreciates the hospitality of the Universidad de Guanajuato.

\vfill
\newpage
\noi { \bf REFERENCES}

\begin{itemize}
\item[1.-] M. Gell-Mann, D. Sharp and W. Wagner, Phys. Rev. Lett. {\bf 8}, 261 (1962).

\item[2.-] See {\it e.g.} P. Rotelli and M.D. Scadron, Nuovo Cimento {\bf A 15}, 643 (1973).

\item[3.-] R.L. Thews, Phys. Rev. {\bf D 10}, 2993 (1974).

\item[4.-] V.L. Eletsky, B.L. Isffe and Y.I. Kogan, Phys. Lett.  
           {\bf B 122},  423 (1983); L.I. Reinders, H.R. Rubinstein and S. Yazaki,
           Nucl. Phys. {\bf B 213}, 109 (1983); S. Narison and N. Paver, Z. 
           Phys. {\bf C 22}, 69 (1984).

\item[5.-] M. Lublinsky, Phys. Rev. {\bf D 55} , 249 (1996). 
                                      
\item[6.-] V.M. Braun, I.E. Filyanor, Z. Phys. {\bf C 44},  157 (1989).

\item[7.-] E.A. Kuraev and Z.K. Silagdze, Phys. Atom. Nucl. {\bf 58}, 1589 (1995).

\item[8.-] Particle Data Group: R.M. Barnett {\it et al}, Phys. Rev.  {\bf D 54}, 
           Part. I,  1 (1996).

\item[9.-] G. L\'opez Castro and D. A. L\'opez Falc\'on, Phys. Rev. {\bf D 54},  4400
(1996).

\item[10.-] See {\it e.g.} Eqs. (3,4,9) of P. Ko, J.Lee and H. S. Song,  Phys. Lett. 
           {\bf B 366}, 287 (1996).

\end{itemize} 
\newpage
{\bf Figure captions:}

\bi
\vskip2ex
\centerline{
\epsfxsize=400pt
\epsfbox{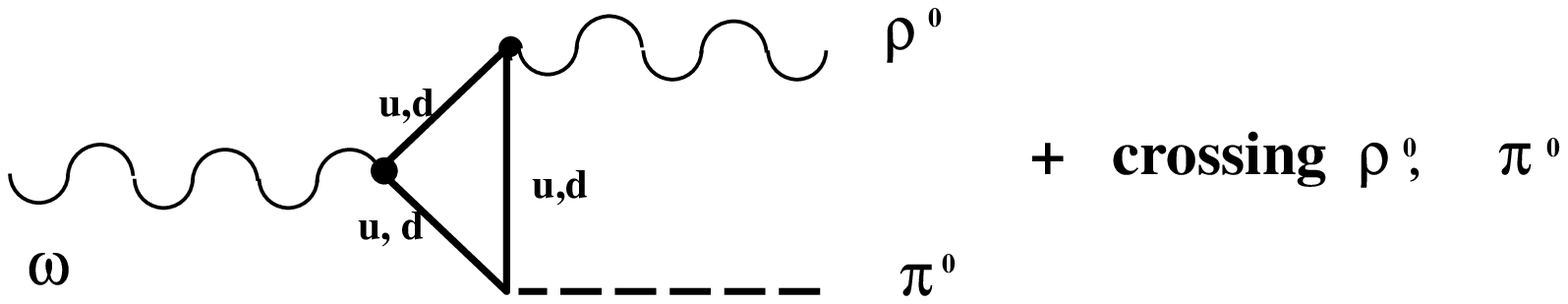}}
\vskip4ex
\bc
{\small{Fig. 1}\\
 Quark loop triangles contribution for $\omega \to \rho^0 \pi^0$ transition. }
\ec

\bi

\vskip2ex
\centerline{
\epsfysize=150pt
\epsfbox{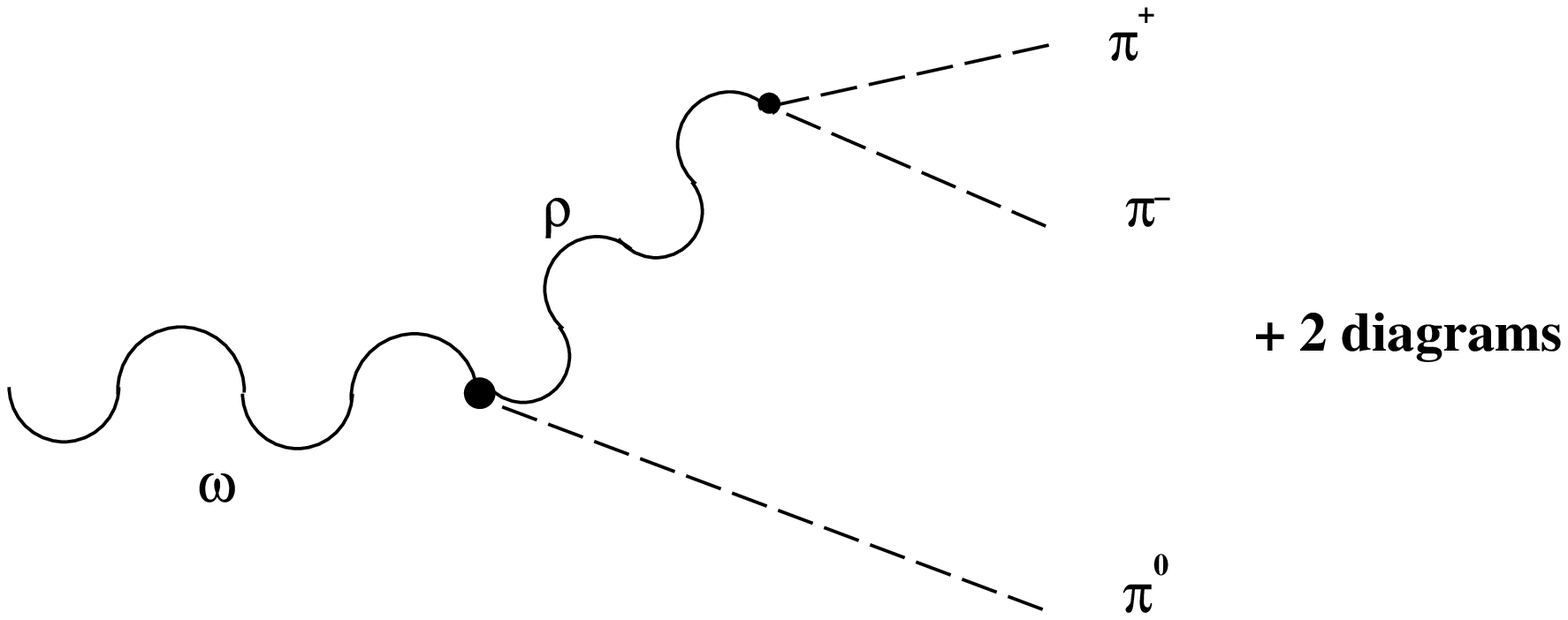}}
\vskip4ex
\bc
{\small{Fig. 2}\\
Intermediate $\rho$ contributions for the $\omega  \to \pi^0 \pi^+\pi^-$ 
decay.}
\ec

\bi

\vskip2ex
\centerline{
\epsfxsize=400pt
\epsfbox{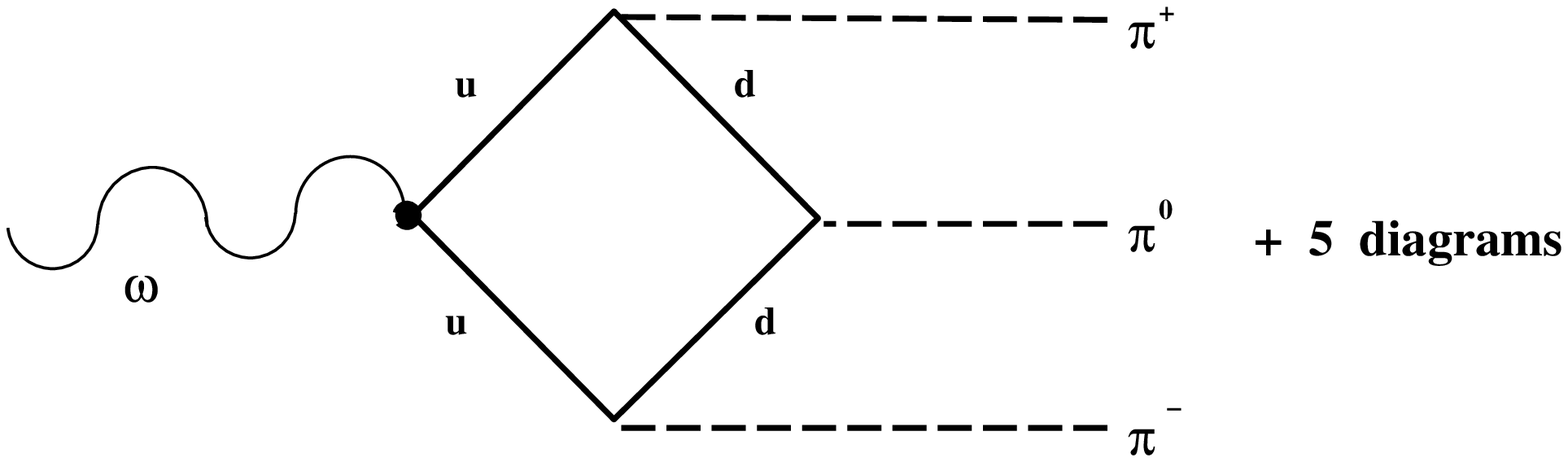}}
\vskip4ex
\bc
{\small{Fig. 3}\\
Quark boxes for $\omega \to \pi^+ \pi^0 \pi^-$ decay.}
\ec

\bi
\noindent

\end{document}